\documentclass[12pt]{article}
\usepackage{epsfig,amsfonts,amsthm}
\usepackage{latexsym}
\usepackage{amsmath}
\usepackage{amssymb}
\usepackage{bm}
\newcommand{\be}{\begin{equation}}
\newcommand{\ee}{\end{equation}}
\newcommand{\bea}{\begin{eqnarray}}
\newcommand{\eea}{\end{eqnarray}}
\newcommand{\dst}{\displaystyle}
\newcommand{\fr}[2]{\frac{{\dst #1}}{{\dst #2}}}
\newcommand{\br}{{\bf r }}

\newcommand{\bk}{{\bf k}}
\newcommand{\bK}{{\bf K}}

\renewcommand{\Re}{\mbox{Re}}
\renewcommand{\Im}{\mbox{Im}}
\newcommand{\lr}[1]{ \langle #1 \rangle}

\def\lsim{\mathrel{\rlap{\lower4pt\hbox{\hskip1pt$\sim$}}
    \raise1pt\hbox{$<$}}}         

\def\gsim{\mathrel{\rlap{\lower4pt\hbox{\hskip1pt$\sim$}}
    \raise1pt\hbox{$>$}}}         

\title{Measuring the phase of the scattering amplitude with vortex beams}
\author{Igor~P.~Ivanov
\\
  {\small IFPA, Universit\'{e} de Li\`{e}ge, All\'{e}e du 6 Ao\^{u}t 17, b\^{a}timent B5a, 4000 Li\`{e}ge, Belgium}\\
  {\small and}\\
  {\small Sobolev Institute of Mathematics, Koptyug avenue 4, 630090, Novosibirsk, Russia}\\
  }

\begin{document}

\maketitle

\begin{abstract}
We show that colliding vortex beams instead of (approximate) plane waves
can lead to a direct measurement of how the overall phase of the plane wave scattering amplitude 
changes with the scattering angle. 
Since vortex beams are coherent superpositions of plane waves with different momenta,
their scattering amplitude receives contributions from plane wave amplitudes with
distinct kinematics. 
These contributions interfere, leading to the measurement of their phase difference.
Although interference exists for any generic wave packet collision,
we show that using vortex beams dramatically enhances sensitivity to the phase 
in comparison with non-vortex beams.
Since the overall phase is inaccessible in a plane wave collision,
this measurement would be of great importance for a number of topics in hadronic physics,
for example, meson production in the resonance region, physics of nucleon resonances,
and small angle elastic hadron scattering.
\end{abstract}

\section{Introduction}

\subsection{The phase of the scattering amplitude}

Scattering of energetic particles on targets and on each other is the standard way 
to investigate their structure and interactions.
Strength of a scattering process is characterized by the complex 
scattering amplitude ${\cal M} = |{\cal M}| e^{i\alpha}$. 
If we have in mind the usual situation of plane-wave scattering 
(initial and final state momenta are generically denoted as $k_i$ and $k_f$, respectively),
then it is only the absolute value of the amplitude $|{\cal M}(k_f,k_i)|$ which is experimentally accessible
via the cross section measurement, but not the overall phase $\alpha(k_f,k_i)$.
 

Despite not being directly measurable in plane wave scattering,
$\alpha(k_f,k_i)$ is still a well-defined quantity which can be traced back to partial wave phase shifts.
This phase can be of much importance, especially in hadronic physics, where a large phase is usually a sign of 
complicated dynamics. For example,
in hadronic production or scattering experiments in the resonance region the phase of the amplitude traces a characteristic
curve as a function of the invariant mass when one sweeps across a resonance.
Another example is the high-energy elastic scattering of hadrons (notably, $pp$ and $p\bar p$).
Due to the dynamical properties of the $t$-channel exchange with the vacuum quantum numbers,
dubbed the Pomeron, (the hadronic part of) the scattering amplitude becomes mostly imaginary,
but it also contains a real part. Their ratio, usually quantified by the parameter $\rho(s,t) = \Re {\cal M}/ \Im {\cal M}$,
turns out to be a sensitive probe of the Pomeron models, see for example \cite{PomeronModels} and references therein.

Being so important for hadronic physics, the phase of the scattering amplitude is impossible 
or notoriously difficult to measure in the usual (approximate) plane wave scattering.
Only when the reaction can proceed via the hadronic and the electromagnetic mechanisms
and when the electromagnetic phase is sufficiently constrained,
can the phase of the hadronic amplitude be accessed via the interference pattern between the two amplitudes.
However, even this kind of measurement often poses experimental challenges. 
For example, in high-energy elastic processes the phase of the hadronic amplitude can be measured
via the Coulomb-hadronic interference in a very small-$t$ domain, \cite{WY}, where the phase of the Coulomb amplitude
itself is a matter of debates, see \cite{Blois2009}.
As a result, the $\rho$-parameter remains poorly constrained at high energies even for the $pp$/$p\bar p$ scattering, 
let alone other hadrons. 
The phase of the Pomeron amplitude in diffractive photo- or electroproduction of $\pi^+\pi^-$ pairs
can be measured via its interference with the Primakoff mechanism, \cite{PomeronPhase},
but this suggestion has not yet been tested experimentally.

In other cases, when there is no clear reference mechanism with a known phase available,
such as process $\gamma p \to K^+\Lambda$,
the overall phase determination becomes impossible, \cite{photoproduction}.
This represents the notorious overall phase ambiguity in the interpretation of meson photoproduction data
and the spectroscopy of baryonic resonances. 
Indeed, disentangling several interfering partial waves (and therefore the possible baryonic resonances in these channels) 
depends on how the phase changes with the polar angle of the produced meson. 

\subsection{The essence of the proposal}

Here we propose a novel type of experiment which can probe 
the phase of the plane wave scattering amplitude $\alpha(k_f,k_i)$.
To be more accurate, 
it measures how this phase changes with the production angle
(its absolute magnitude still remains unmeasurable).
This phase sensitivity comes from the fact that the scattering amplitude 
of two collinear Bessel vortex beams (to be defined below) is a coherent sum of two plane-wave amplitudes
corresponding to fixed final momenta $k_f$ but slightly different initial momenta, $k_i$ and $k'_i$.
These plane-wave amplitudes interfere, and as the result one gains sensitivity to their relative
phase $\alpha(k_f,k_i)-\alpha(k_f,k'_i)$.

In principle, interference among several plane wave amplitudes arises in scattering 
of any generic wave packets.
The crucial feature of our suggestion is to exploit a very special form of transverse wave packets:
the so-called {\em vortex beams}, in which wave fronts have helical shape
and contain phase vortices.
We will show that the phase sensitivity of the vortex beam collision is dramatically better than
for any generic (for example, simple Gaussian) non-vortex wave packets.

In order to make the proposed idea even more transparent, let us compare it with 
the method of partial waves, in which the initial and final states are expanded 
in spherical waves with the 
given orbital angular momentum $\ell$ and its projection $m$.
The phase $\alpha_\ell$ of the partial wave scattering amplitude is defined by its (real) phase shift $\delta_\ell$
and the inelasticity of the scattering.
Since the plane wave is a superposition of many partial waves,
measurement of the full $4\pi$ angular dependence of the plane wave cross section can reveal the 
phase differences $\alpha_\ell - \alpha_{\ell'}$, but cannot access 
the overall plane wave phase $\alpha(k_f,k_i)$.
On the other hand, if we were able to prepare and collide initial states as pure partial waves with given $\ell$,
which would be superpositions of plane waves coming from all directions,
we could measure individual cross sections $\sigma_\ell$ and deduce the angular dependence 
of the plane wave phase $\alpha(k_f,k_i)$.
The problem, of course, is that preparation of initial states coherent over the entire solid angle is unfeasible.

A vortex beam is a cylindrical wave, which represents an intermediate case
between a plane wave and a spherical harmonic. It is a superposition of plane waves
with different but close momenta, and therefore vortex beam scattering is also sensitive to the relative phases of plane wave amplitudes.
In contrast to spherical waves, however, vortex beams {\em can} be created experimentally.

\subsection{Vortex beams}

Optical vortex beams, that is, laser beams carrying non-zero orbital angular momentum (OAM) with respect to the beam axis,
are well-known and routinely used in optics, \cite{OAM,OAMreview}.
Their application range from microscopy to astrophysics, \cite{TP}. 
Wavefronts of such a beam are not planes but helices, and each photon in this light field (a {\em twisted photon}) carries 
a well-defined OAM quantized in units of $\hbar$. Vortex beams can exist for massive particles as well.
Recently, following the suggestion of \cite{bliokh2007}, electron beams carrying OAM were experimentally demonstrated,
first using the phase plates \cite{twisted-electron} and then with the aid of diffraction gratings \cite{twisted-electron2}.
Such electrons carried the energy as high as 300 keV and the orbital quantum number up to $m\sim 100$.
These vortex beams can be manipulated and focused just as the usual electron beams,
and very recently remarkable focusing of a vortex electron beam to the focal spot of less than 1.2\AA\, in diameter 
was achieved \cite{focusing}.

It is conceivable that future progress will lead to creation of even more energetic twisted particles.
For example, it was recently shown that
X-ray beams carrying orbital angular momentum can be generated in ondulators and 
X-ray free-electron lasers, \cite{Xrayproposal},
and there exist plans to study them at the Argonne National Laboratory, \cite{Xrayproposal2}.
Besides, it was noted in \cite{serbo} that Compton
backscattering of twisted optical photons off an
ultra-relativistic electron beam generates high-energy photons
carrying non-zero OAM. Since the technique of
Compton backscattering is well established \cite{backscattering} and is actively used in nuclear and 
particle physics (for example, at the SPring-8, \cite{backscatteringuses}, and HIgammaS, \cite{backscatteringuses2}, facilities),
realization of this proposal seems feasible with today's technology. 

Vortex beams of other particles, for example of protons, can also be created experimentally.
For low and intermediate energies one could use the same fork diffraction grating technique
as used for electrons.
Alternatively, one can exploit the vortex transfer which spontaneously happens 
in an elastic collision of a lower energy electron vortex beam with a high-energy proton beam 
that initially carried no phase vortex, \cite{entanglement}. 

Although these proposals have not yet been tested experimentally, 
one can start thinking about what new insights into properties of particles, 
and especially hadrons, one can gain in collisions of energetic vortex beams.
First studies along these lines were conducted in \cite{ivanov2011},
where it was noted there that such processes can probe observables which are not accessible
in the plane wave collisions.
The phase measurement discussed in this paper represents 
one particular example of such observables.

The structure of the paper is the following.
In Section~\ref{section-describing} we introduce the formalism of Bessel-type twisted states.
Then, in Section~\ref{section-scattering} we develop further the theory of 
collision of two twisted particles and obtain the expression for its cross section.
Section~\ref{section-phase-angular} explains how one can measure the angular dependence of the
phase of the amplitude.
In Section~\ref{section-discussion} we discuss what exactly is needed to actually 
realize the measurements we propose.
Finally, in the last Section we draw our conclusions.

\section{Describing twisted states}\label{section-describing}

Here we briefly summarize the formalism of Bessel vortex states, 
which are the simplest states carrying a phase vortex. 
We focus on the scalar case and use the notation of \cite{serbo}.
We will accompany three-dimensional vectors with arrows,
while the transverse vectors will be given in bold.

A Bessel state is a non-plane wave solution of the free wave equation 
with a definite frequency $\omega$, longitudinal momentum $k_z$, 
modulus of the transverse momentum $|\bk|=\varkappa$ 
and a definite $z$-projection of orbital angular momentum (orbital helicity) $m$.
When written in cylindric coordinates $r, \varphi_r, z$, this solution
$|\varkappa,m\rangle$  has the form
 \be
|\varkappa, m\rangle = e^{-i\omega t + i k_z z} \cdot
\psi_{\varkappa m}(\br)\,, \quad \psi_{\varkappa m}(\br) = {e^{i m
\phi_r} \over\sqrt{2\pi}}\sqrt{\varkappa}J_{m}(\varkappa r)\,,
 \label{twisted-coordinate}
  \ee
where $J_m(x)$ is the Bessel function. 
The transverse spatial distribution is normalized according to
 \be
\int d^2\br\, \psi^*_{\varkappa' m'}(\br)\psi_{\varkappa m}(\br) = \delta_{m m'}\,
\delta(\varkappa-\varkappa')\,.
 \ee
The azimuthal angle dependence $\propto e^{im\phi_r}$ is the hallmark of the phase vortex.
A twisted state can be represented as a superposition of plane
waves:
 \be
|\varkappa,m\rangle = e^{-i\omega t + i k_z z} \int {d^2\bk
\over(2\pi)^2}a_{\varkappa m}(\bk) e^{i\bk \br}\,,
 \label{twisted-def}
  \ee
where
 \be
a_{\varkappa m}(\bk)= (-i)^m
e^{im\phi_k}\sqrt{2\pi}\;{\delta(|\bk|-\varkappa)\over
\sqrt{\varkappa}}\,.\label{a}
  \ee
This expansion can be inverted:
 \be
e^{-i\omega t + i k_z z} \cdot e^{i\bk\br} =  \sqrt{2\pi
\over\varkappa} \sum_{m=-\infty}^{+\infty} i^m e^{-im\varphi_k}
|\varkappa,m\rangle \,,\quad \varkappa = |\bk|\,. \label{PW}
 \ee
This invertible linear map between plane waves and twisted states means that twisted states simply represent 
another complete basis for transverse wave functions.
When switching to vortex beams, we do not lose any asymptotic state we had before
nor introduce any new state.
More properties of twisted states can be found in
\cite{serbo,ivanov2011}. Here we only note that Eq.~(\ref{twisted-def}) in fact describes the passage
from plane waves to twisted particles in description of a scattering process: 
one just needs to apply at the level of the scattering matrix 
the $a_{\varkappa m}$- or $a^*_{\varkappa m}$-weighted transverse momentum
integration for each incoming or outgoing twisted particle.

A pure Bessel state $|\varkappa,m\rangle$ with fixed $\varkappa$ is non-normalizable in the transverse plane.
A much better approximation to physically realizable states such as Bessel-Gaussian or aperture-limited beams
is given by a fixed-$m$ superposition of Bessel states 
\be
|\varkappa_0,\sigma;m\rangle = \int d\varkappa \, f(\varkappa) |\varkappa,m\rangle\,,\label{WP}
\ee
with a properly normalized weight function $f(\varkappa)$ peaked at $\varkappa_0$ and 
having width $\sigma$. This state, which we call the transverse Bessel wave packet, 
is normalizable (and localized) in the transverse plane.
For calculational simplicity we assume that such states are monochromatic,
that is, $k_z$ is supposed to vary with $\varkappa$ so that $k_z^2 + \varkappa^2$ is constant.
Properties of such states were discussed in \cite{IS2011}. 
Here we note that they play the key role in resolving the non-forward-to-forward transition paradox 
in Bessel beam scattering which arose from contradicting results of \cite{ivanov2011} and \cite{serbo}.

When discussing twisted states of fields with polarization degrees of freedom, extra care should be taken.
For example, in the case of photon the issues arising span from the absence of a well-defined polarization vector, which is replaced by 
the polarization field, to the problem of gauge-invariant separation of spin and OAM operators.
Fortunately, these problems can be avoided in the paraxial approximation, namely at $\varkappa \ll \omega$,
where the polarization and OAM degrees of freedom can be reliably separated, see discussion in \cite{ivanov-talk-2011}.
The polarization field can be then approximated by a constant polarization vector lying in the plane orthogonal
to the axis $z$. The case of twisted fermions, for which the paraxial approximation is also assumed, can be treated similarly.
A detailed account of the ultrarelativistic twisted electron beyond the paraxial approximation 
was recently given in \cite{bliokh2011}.

\section{Beam-beam scattering beyond the plane wave approximation}\label{section-scattering}

\subsection{General description}

The general theory of scattering of arbitrarily shaped beams was developed in \cite{theoryMD}.
Its formulation took into account not only the arbitrary spatial distribution, but also a non-uniform temporal dependence
of the non-monochromatic beams. The basic objects describing the beams were not the coordinate or momentum wave functions,
but the Wigner distributions which arose after statistical averaging over the ensemble of particles in each beam.
This formalism was especially useful for description of coherent processes such as beamstrahlung, in which a particle
from one beam coherently scatters from many particles in the counterpropagating beam.

For the purpose of the present paper (namely, to demonstrate sensitivity of the vortex beam scattering to the phase of the amplitude) 
such a general set-up is not needed. Here, we make the following assumptions: (1) the colliding particles 
are in pure states describable by wave functions, (2) the beams are monochromatic. 
These two assumptions are made solely to simplify the calculations;
if needed, one can redo the analysis in the most general framework of \cite{theoryMD}.
Finally, at a later stage we will also use the fact that the colliding beams are paraxial, and the longitudinal distribution
of their envelopes is approximately uniform over a long distance compared to the transverse size. 

We consider a $2 \to 2$ scattering (not necessarily elastic) 
and start with the plane-wave collision when initial particles with four-momenta $k_1$ and $k_2$
scatter into final particles with momenta $k'_1$ and $k'_2$. The total momentum
is denoted as $K = k_1 + k_2 = k'_1 + k'_2$. The $S$-matrix element of this process is
\be
S_{PW} = i(2\pi)^4\delta^{(4)}(k_1 + k_2 - k'_1 -k'_2) {{\cal M}(k_1,k_2) \over \sqrt{16 E_1 E_2 E'_1 E'_2}}\,, \label{SPW}
\ee
where the invariant amplitude ${\cal M}$ is calculated by the standard Feynman rules.
This amplitude, in general, depends on all the momenta involved, but for future convenience we explicitly indicate 
only the initial momenta.

Let us now assume that the initial particles are not plane waves but are represented by coordinate wave functions
$\psi_1(\vec r)$ and $\psi_2(\vec r)$ normalized by $\int d^3 r |\psi_i(\vec r)|^2 = 1$. The corresponding momentum-space 
wave functions are
\be
\varphi(\vec k) = \int d^3 r\, \psi(\vec r)\, e^{-i \vec k \vec r}\,,\quad
\int {d^3 k \over (2\pi)^3} |\varphi(\vec k)|^2 = 1\,.
\ee
If these wave functions are not normalizable in the entire space, then a large but finite integration volume is assumed 
in order to provide the regularization.

The $S$-matrix element for scattering of this initial state into the same plane-wave final state with momenta $k'_1$ and $k'_2$,
can be written as
\be
S = \int {d^3 k_1 \over (2\pi)^3} {d^3 k_2 \over (2\pi)^3} \varphi_1(k_1) \varphi_2(k_2) S_{PW}\,.
\ee
Since the beams are monochromatic,
the number of scattering events into a given differential volume of the final phase space per unit time is
\be
d\nu = {(2\pi)^7 \delta(E) \over 4 E_1 E_2} \, |F|^2 \, {d^3 k'_1 \over (2\pi)^3 2E'_1} {d^3 k'_2 \over (2\pi)^3 2E'_2}\,.\label{dnu1}
\ee
Here $\delta(E)$ stands for $\delta(E_1+E_2-E'_1-E'_2)$, and
\be
F =  \int {d^3 k_1 \over (2\pi)^3} {d^3 k_2 \over (2\pi)^3}  \varphi_1(k_1) \varphi_2(k_2) \delta^{(3)}(\vec k_1 + \vec k_2 - \vec K)\, {\cal M}(k_1,k_2) \,.
\ee

Let us now discuss the obvious but important issue of the final momentum distribution of the produced particles.
In the conventional plane-wave case the final momenta are constrained by the overall momentum conservation.
The value of $\vec K$ is then fixed, and the momenta $\vec k'_1$ and $\vec k'_2$ become maximally correlated.
The usual procedure then is to use the momentum delta-function and remove, for example, the $\vec k'_2$ integration. 
In our case, (\ref{dnu1}), this maximal correlation is absent. One must consider distribution of the events
over momenta $\vec k'_1$ and $\vec k'_2$, or alternatively, over $\vec k'_1$ and $\vec K$,
where the exact $\vec K$ region depends on $\varphi_i(\vec k_i)$.

\subsection{Approaching the plane wave limit}

Before proceeding further with vortex beams, let us first see how (\ref{dnu1}) simplifies in the plane-wave limit. 
The plane-wave limit corresponds to very compact momentum wave functions $\varphi_i(\vec k_i)$ localized
near $\lr{\vec k_i}$. The matrix element can then be approximated as
${\cal M}(k_1, k_2) \approx {\cal M}(\lr{k_1}, \lr{k_2}) \equiv {\cal M}_0$, and the expression for $f$ becomes
\be
F  = {\cal M}_0 \int {d^3 r \over (2\pi)^3} e^{-i\vec K \vec r} \psi_1(r)\psi_2(r)\,.
\ee
Changing $d^3 k'_1 d^3 k'_2$ to $d^3 k'_1 d^3 K$ and integrating $|f|^2$ over $\vec K$, one gets
\be
\int d^3 K |f|^2 =  |{\cal M}_0|^2 \int {d^3 r\over (2\pi)^3} |\psi_1(\vec r)|^2  |\psi_2(\vec r)|^2\,.
\ee
This means that in the plane wave limit, $k_i = \lr{k_i}$, one can effectively replace
\be
|F|^2 \to  |{\cal M}_0|^2  \delta^{(3)}(k_1 + k_2 - K) \int {d^3 r\over (2\pi)^3} |\psi_1(\vec r)|^2  |\psi_2(\vec r)|^2\,.
\ee
The number of events can therefore be split into the cross section and luminosity factors:
\bea
&& d\nu = d\sigma \cdot L\,,\label{dnuPW}\\[2mm]
&& d\sigma = {(2\pi)^4 \delta^{4}(k_1+k_2-k'_1-k'_2) \over 4 E_1 E_2 v } \, |{\cal M}_0|^2 
{d^3 k'_1 \over (2\pi)^3 2E'_1} {d^3 k'_2 \over (2\pi)^3 2E'_2}\,,\nonumber\\ 
&& L = v \int d^3 r\, n_1(\vec r) n_2(\vec r)\,,\quad n_i(\vec r) \equiv |\psi_i(\vec r)|^2\,.\label{lumiPW}
\eea
Note that we inserted here by hand the relative velocity of the two plane waves, $v = |\vec v_1 - \vec v_2|$.

It must be stressed that separation of the number of events into the differential cross section and 
the (conventional) luminosity is uniquely defined only for plane waves.
Extending this splitting for non-plane-wave collisions is a matter of convention. In this way, one needs
to (somewhat arbitrarily) introduce the notion of {\em generalized cross section}, \cite{theoryMD}, for example,
by dividing the full $d\nu$ in (\ref{dnu1}) by $L$ (\ref{lumiPW}) with $v$ defined for $\lr{\vec k_i}$
rather than $\vec k_i$.
With this definition, the generalized cross section for non-plane-wave scattering
takes form
\be
d\sigma = d\sigma_0\, R\, d^3 K \,,\quad R \equiv {(2\pi)^3\, |F|^2 \over |{\cal M}_0|^2  \int d^3 r\, n_1(\vec r) n_2(\vec r)}\,,
\label{ratioR1}
\ee
where $d\sigma_0$ is the plane-wave $\vec K$-integrated cross section.
In this notation, the plane-wave limit corresponds to $R \to  \delta^{(3)}(k_1 + k_2 - K)$.

\subsection{Pure Bessel states}

Let us now consider the particular case of pure Bessel states for both initial particles.
This problem was first considered in \cite{ivanov2011},
where this process was termed the ``double-twisted scattering'' and 
a generic expression for the cross section was found. That expression used a definition
of the flux factor (i.e. luminosity) different from what was described above, 
therefore that cross section differs in the overall normalization from our results.
However, the main finding remains unchanged: 
for each $\vec K$ there exist two kinematic configurations which add up coherently.
Since this finding is the key to our proposal, we discuss it below in more detail.

Consider a collision of two Bessel states $|\varkappa_1,m_1\rangle$ and $|\varkappa_2,m_2\rangle$.
For simplicity we assume that both vortex states are coaxial, that is, they are defined
with respect to the same axis $z$. The wave functions describing these states 
are given by regularized versions of (\ref{twisted-coordinate}) and (\ref{twisted-def}).
Namely, we assume a large but finite cylindrical quantization volume with macroscopic length $L_z$
and radius $R$.
The coordinate wave functions are then
\be
\psi_1(\vec r) = c\, e^{ik_{1z}z} e^{im_1\phi_{r}}\sqrt{{\varkappa_1 \over 2\pi}}J_{m_1}(\varkappa_1 r)\,,
\quad
\psi_2(\vec r) = c\, e^{ik_{2z}z} e^{-im_2\phi_{r}}\sqrt{{\varkappa_2 \over 2\pi}}J_{-m_2}(\varkappa_2 r)\,,
\ee
where the orbital helicity $m_2$ is accompanied by the sign minus  because the second particle moves in the negative $z$ direction.
The normalization coefficient $c$ is fixed by
\be
1 = |c|^2 L_z \varkappa_i \int_0^R rdr\, J^2_{m_i}(\varkappa_i r)  \approx |c|^2 {L_z R \over \pi}\,,
\ee
where we used the large-argument asymptotics for the Bessel function.
The wave function in the momentum representation, for example, of the first particle is
\be
\varphi_1(k) = 2\pi c\, \delta_{r.}(k_z - k_{1z})\cdot (-i)^m e^{im_1 \phi_{k}} \sqrt{{2\pi \over \varkappa_1}}\delta_{r.}(|\bk|-\varkappa_1)\,,
\ee
where two $\delta_{r.}$-functions denote regularized versions of the delta-function. 
For the longitudinal/transverse dynamics, $\delta_{r.}(q)$ is localized within $|q| < 1/L_z$ and $|q| < 1/R$, respectively, 
and its integral over all $q$ gives unity.
The longitudinal factors can be easily accounted for, and the cross section (\ref{ratioR1}) can now be represented as
\be
{d\sigma \over d^2\bK} = d\sigma_0\cdot R_\perp \,,\quad \mbox{where}\quad R_\perp = 
{(2\pi)^2\, |F_\perp|^2 \over |{\cal M}_0|^2  \int d^2 \br\, n_{1}(\br) n_{2}(\br)}\,,\label{Rperp}
\ee
where
\be
F_\perp = \int {d^2 \bk_1 \over (2\pi)^2} {d^2 \bk_2 \over (2\pi)^2} a_{\varkappa_1,m_1}(\bk_1) a_{\varkappa_2,-m_2}(\bk_2) 
\delta^{(2)}(\bk_1 + \bk_2 - \bK)\, {\cal M}\,.\label{fperp}
\ee
Although we refer here to $a_{\varkappa, m}(\bk)$ defined by (\ref{a}), it is assumed that the delta-function inside it
is the regularized one.

\begin{figure}[!htb]
   \centering
\includegraphics[width=12cm]{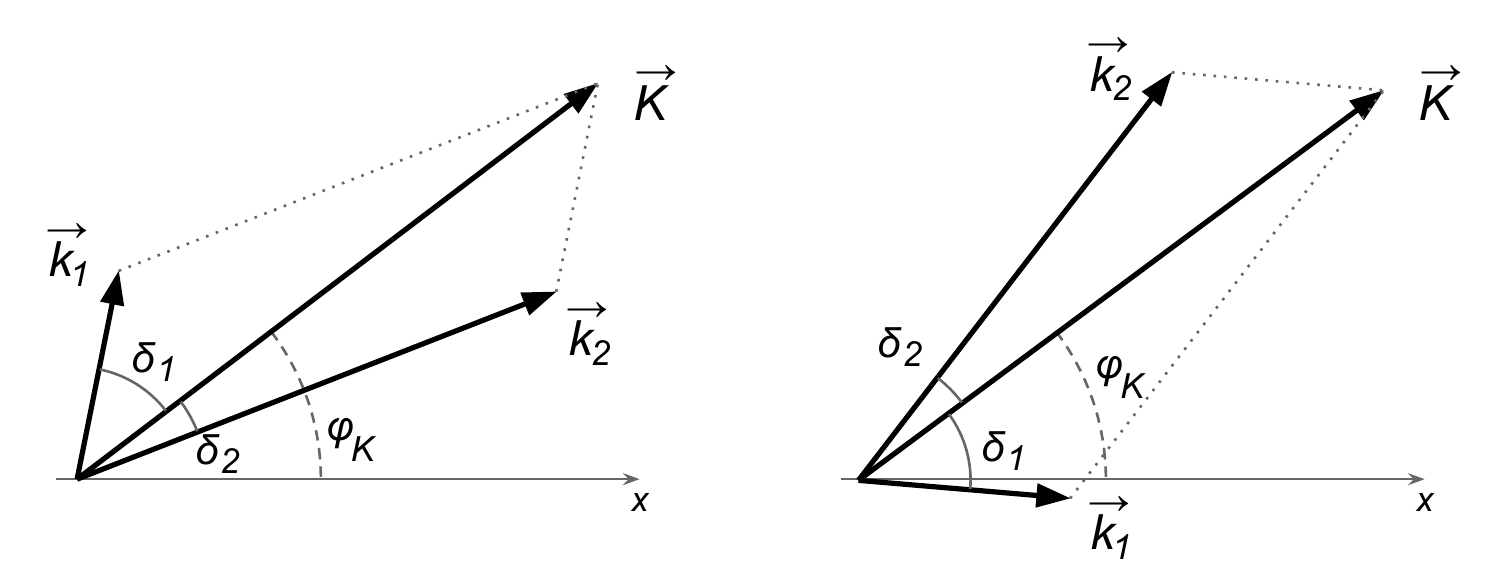}
\caption{Two interfering kinematical configurations on the transverse plane for vectors $\bk_1$ and $\bk_2$ of fixed length 
which sum up to the same vector $\bK$.}
   \label{fig-2-configurations}
\end{figure}

The integrals of this type in the limit $R\to \infty$ were studied in \cite{ivanov2011}.
In this limit, $a_{\varkappa, m}$'s contain true delta-functions, so that all four integrations in (\ref{fperp}) are eliminated.
It turns out that this integral is non-zero, only if there exists a triangle with sides $\varkappa_1$, $\varkappa_2$, $|\bK|$,
so that $\bK$ lies in the ring 
\be
|\varkappa_1 - \varkappa_2| \le |\bK| \le \varkappa_1 + \varkappa_2\,.\label{Kregion}
\ee
For any $\bK$ lying in this ring there exist exactly two kinematic configurations shown in Fig.~\ref{fig-2-configurations} for $\bk_1$ and $\bk_2$ with azimuthal angles
\be
\phi_1 = \phi_K \pm \delta_1\,, \quad \phi_2 = \phi_K \pm (-\delta_2)\,,\quad
\delta_{1,2} = \arccos\left({\varkappa_{1,2}^2 + \bK^2 - \varkappa_{2,1}^2 \over 2\varkappa_{1,2} |\bK|} \right)\,.
\label{phi1phi2}
\ee
Note that by definition $0 \le \delta_1 + \delta_2 \le \pi$.
The momenta  $\bk_1$ and $\bk_2$ corresponding to the sign $+$ in both $\pm$'s are denoted as $\bk_1^{+}$ and $\bk_2^{+}$
(Fig.~\ref{fig-2-configurations}, left),
while for the sign minus they are denoted as $\bk_1^{-}$ and $\bk_2^{-}$ (Fig.~\ref{fig-2-configurations}, right).
The amplitude ${\cal M}$ taken at these two pairs of momenta will be denoted by ${\cal M}_+$ and ${\cal M}_-$, respectively.
With this notation, the integral $F_\perp$ can be written as
\be
F_\perp = (-i)^{m_1-m_2} e^{i(m_1-m_2)\phi_K} {\sqrt{\varkappa_1\varkappa_2} \over (2\pi)^3}\, {1 \over 2\Delta}
\left(e^{im_1\delta_1 + im_2\delta_2}{\cal M}_+ + e^{-im_1\delta_1 - im_2\delta_2}{\cal M}_-\right)\,.
\ee
Here
\be
\Delta = {1 \over 4}\sqrt{2\varkappa_1^2\varkappa_2^2 + 2\varkappa_1^2\bK^2 + 2\varkappa_2^2\bK^2 - \varkappa_1^4 - \varkappa_2^4 - \bK^4} =
{1 \over 2}\varkappa_1 \varkappa_2 \sin(\delta_1 + \delta_2)\,,\label{Delta}
\ee
is the area of the triangle with side $\varkappa_1$, $\varkappa_2$, $|\bK|$.
We then obtain
\be
|F_\perp|^2 = {\varkappa_1 \varkappa_2 \over (2\pi)^6 4\Delta^2}\left[ |{\cal M}_+|^2 + |{\cal M}_-|^2
+ 2\Re\left(e^{2i(m_1\delta_1 + m_2\delta_2)} {\cal M}_+ {\cal M}^*_-\right)\right]\,.\label{fperp2}
\ee

The last term here is the interference term (or the autocorrelation function), which is sensitive 
to the phase difference of the amplitudes ${\cal M}_+$ and ${\cal M}_-$.
It is this term that allows one to measure how the overall phase of the amplitude changes with angles,
which will be the subject of the next Section. 

\subsection{End point singularities}\label{singularities}

It is clear that the expression (\ref{fperp2}) for $|F_\perp|^2$ cannot be integrated, as it stands, in the 
entire $|\bK|$ domain. At the end points $\varkappa_1 = \varkappa_2 \pm |\bK|$, where $\Delta = 0$,
it develops singularities. Indeed, at these points ${\cal M}_+ = {\cal M}_-$,
while $\delta_i = 0$ or $\pi$, so that the expression in brackets has a finite limit.
Changing the integration variable from $\bK^2$ to $\alpha \equiv \delta_1 + \delta_2$,
we get a non-integrable singularity of the type
\be
\int {d\bK^2 \over \Delta^2} \propto \int_0^\pi {d\alpha \over \sin\alpha}\,. \label{singularity}
\ee
This singularity can be traced back to using the $R\to \infty$ limit.
However, in this limit the denominator of $R_\perp$ (\ref{Rperp}) also diverges, and one gets the indeterminate
form $\infty/\infty$. The remedy in this situation is to consider large but finite $R$ and take the limit of the ratio $R_\perp$
rather than its numerator and denominator separately. Then, the exact delta-functions inside $a_{\varkappa m}$'s 
in (\ref{fperp}) are replaced by regularized functions $\delta_{r.}$. The result (\ref{fperp2}) stays valid
until $|\bK|$ comes in the $1/R$ vicinity of the end points, where the function $F_\perp$ gets regularized.
Therefore, at finite $R$ the effective integration limits in (\ref{singularity}) are $\alpha_0$ and $\pi - \alpha_0$
with $\alpha_0 \sim {\cal O}(1/\varkappa R)$, where in the logarithmic accuracy 
$\varkappa$ can be taken equal to any of $\varkappa_i$. 
After integration, the numerator of $R_\perp$ contains a large logarithm $\log(\varkappa R)$.
The denominator of $R_\perp$ (\ref{Rperp}) is also enhanced by a similar logarithm.
Indeed, for large $R$ the integral $\int d^2 n_1(\br) n_2(\br)$ is approximately proportional to
$$
\int^R {dr \over r} \cos^2\left(\varkappa_1 r - {\pi m_1 \over 2} - {\pi \over 4}\right) 
\cos^2\left(\varkappa_2 r + {\pi m_2 \over 2} - {\pi \over 4}\right) \propto \log(\varkappa R)\,.
$$
Therefore, $\int d^2\bk R_\perp$ stays finite in the $R\to \infty$ limit.

Following this analysis, one is tempted to conclude that in the $R \to \infty$ limit the {\em only} contribution
comes from the two end points in the $|\bK|$ region, while the contribution
from the majority of the interval (\ref{Kregion}) is suppressed as $1/\log(\varkappa R)$
and eventually vanishes.
The tricky thing, however, is that this suppression is not so strong.
Taking any reasonable macroscopic $R$ and microscopic $\varkappa$, one gets
$\log(\varkappa R) \sim 50$.
So, the inner region of the $|\bK|$-interval does give a non-vanishing contribution even
for macroscopic quantization volumes.
This is a curious example of a situation when the unphysical limit of infinite quantization volume is quantitatively different 
from the physically motivated extremely large but finite volume. 

The case of a pure Bessel state, although important for the formalism itself, is of academic interest for applications.
Indeed, true Bessel beams of energetic particles with a macroscopic spatial extent cannot be created experimentally. 
Taking into account the vortex beams already created and envisaging the suggestions
for generation of high-energy particles with OAM \cite{serbo,entanglement}, one can count only
on transverse Bessel wave packets (\ref{WP}) with the weight function $f(\varkappa)$ having a reasonably narrow but finite width $\sigma$.
In this case, the ratio $R_\perp$ contains the square of
\be
\tilde F_\perp = \int d\varkappa_1 d\varkappa_2 f_1(\varkappa_1) f_2(\varkappa_2) \, F_\perp
\ee
instead of just $F_\perp$.
In the case $\sigma \ll \varkappa$, one can basically repeat the above analysis using the regularization parameter $1/\sigma$
instead of $R$. In this case, the logarithmic factor $\log(\varkappa/\sigma)$ is not big at all, and the $|\bK|$ integration
gives comparable contributions from the end points and from the central part of the $|\bK|$-region.

\begin{figure}[!htb]
   \centering
\includegraphics[width=8cm]{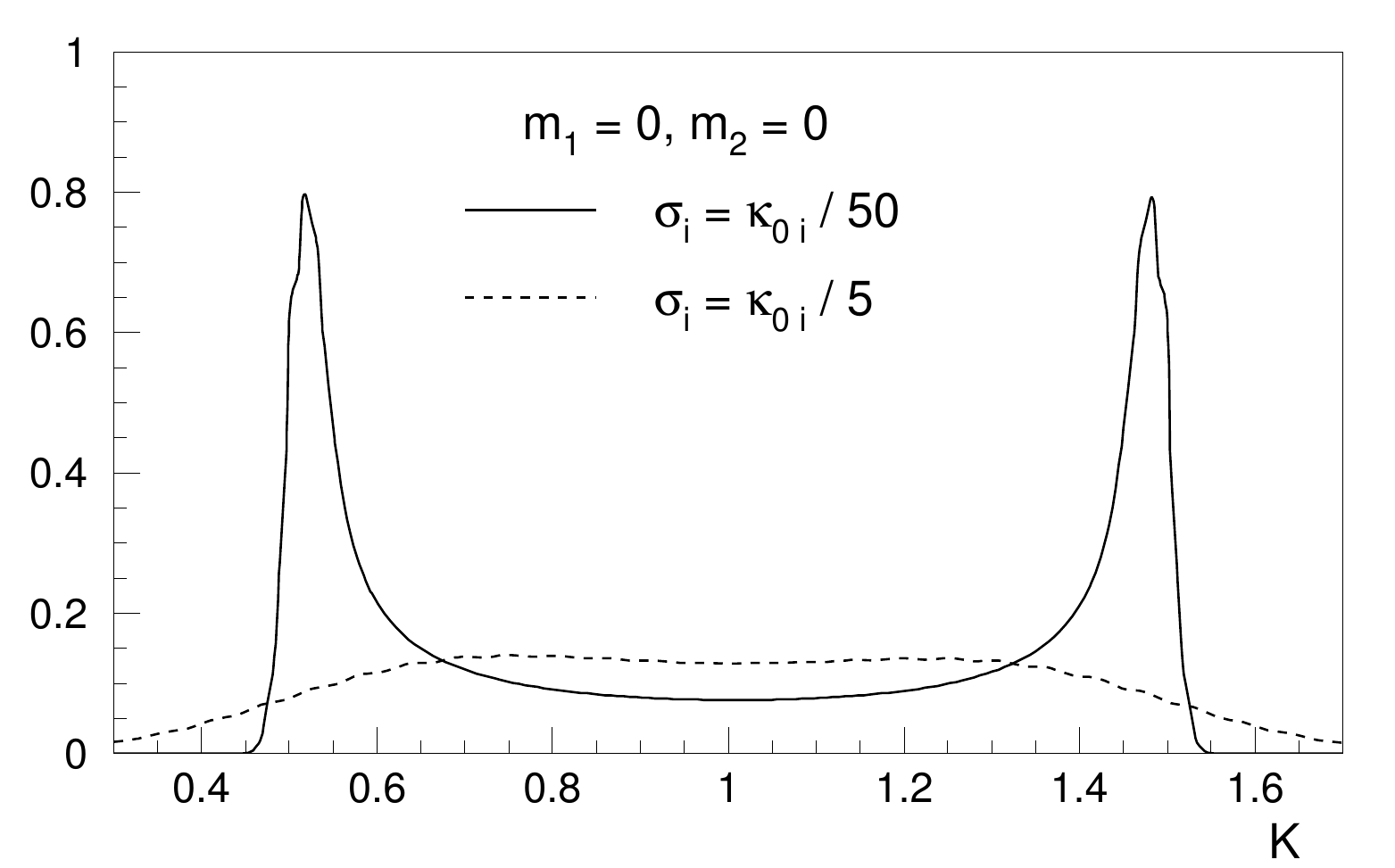}
\hfill
\includegraphics[width=8cm]{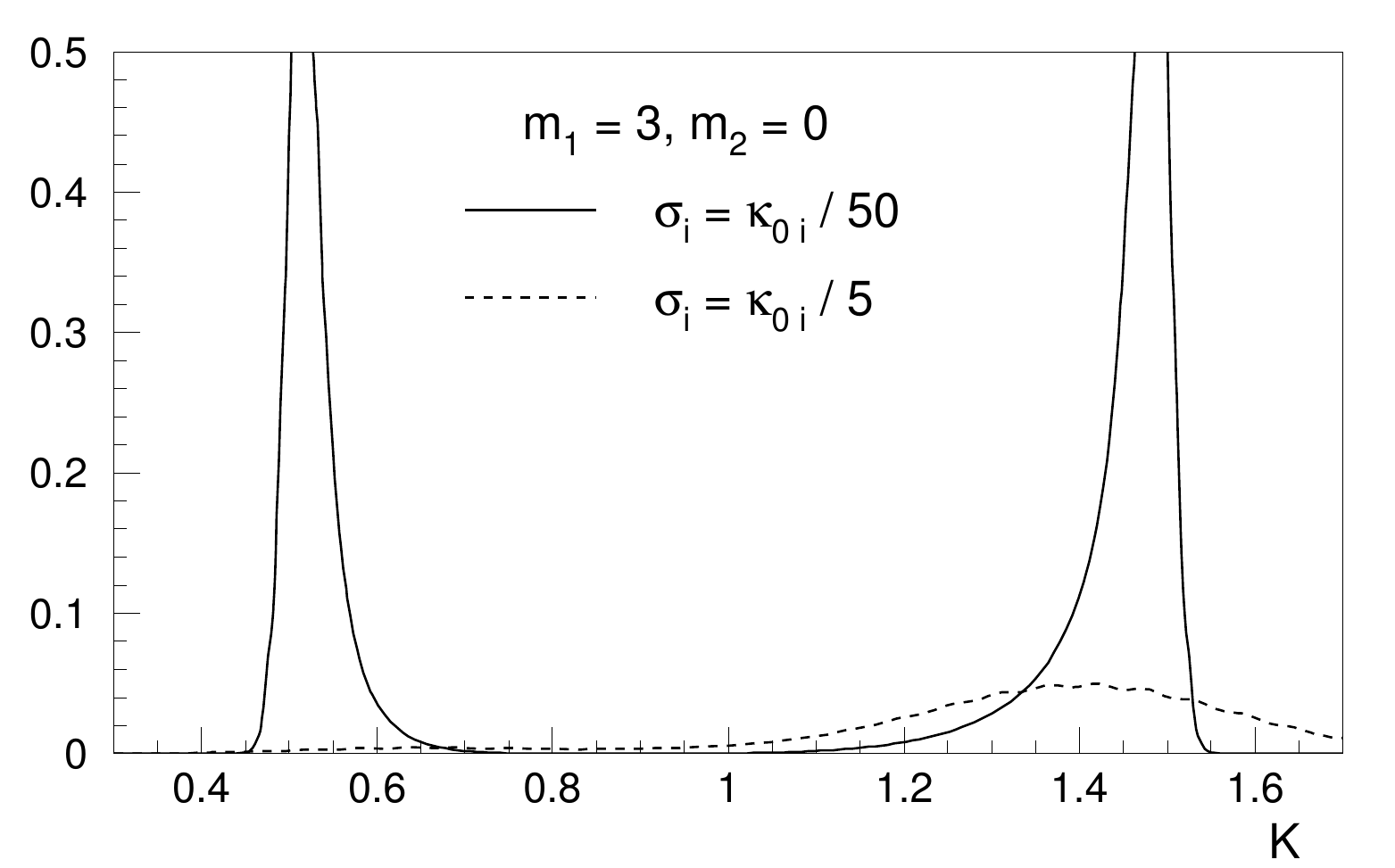}
\caption{Quantity $|\bK|  R_\perp$ as a function of $|\bK|$ (measured in arbitrary units) for $m_1 = 0$ (left pane) and $m_1 = 3$ (right pane);
$m_2 = 0$ in both cases.
The plots were drawn for $\varkappa_1 = 1$ and $\varkappa_2 = 0.5$ measured in the same arbitrary units.
The solid and dashed lines correspond to $\sigma_i = \varkappa_{0i}/50$ and $\sigma_i = \varkappa_{0i}/5$, respectively.
}
   \label{fig-rperp}
\end{figure}

For illustration, we show in Fig.~\ref{fig-rperp} the ratio $R_\perp$ multiplied by $|\bK|$ for the simplest case of constant scattering amplitude: 
${\cal M}_+ = {\cal M}_- = {\cal M}_0$. The qualitative dependences discussed above are well reproduced.
For sufficiently small $\sigma_i$ this quantity indeed peaks near the end points (which are at $0.5$ and $1.5$ in the units chosen),
so that the end points indeed dominate, especially at non-zero orbital helicities.
For not too small $\sigma$, such as $\sigma_i = \varkappa_{0i}/5$, the end points do not dominate over the inner region at all;
however, in this case the non-zero helicities become noticeably suppressed.

\section{Measuring the phase of the amplitude}\label{section-phase-angular}

It was already mentioned above that the expression for the (generalized) cross section in a vortex-beam collision
is sensitive to the difference of the phases of the invariant amplitudes calculated at two sets of the initial momenta.
This sensitivity comes from the interference term in (\ref{fperp2}). In this Section we quantify this sensitivity.

Although the results derived below are rather universal and not particle-type specific, 
we will use for convenience a notation that alludes
to one particularly representative case: the pion photo-production $\gamma p \to \pi p$. 
In this situation, we will refer to the particles with momenta $k_1$
and $k_2$ as the photon and proton, respectively, while $k'_1$ will denote the final pion momentum.

Since the purpose of this paper is to demonstrate the sensitivity to the phase in a generic set-up, 
we will not stick to any particular model for this specific process. 
Instead we consider a toy model, in which the modulus of the amplitude is constant
while the phase depends only on the angle between the photon and the pion in the proton rest frame, $\theta^*_{\gamma \pi}$:
\be
{\cal M} = |{\cal M}|e^{i\alpha(\theta^*_{\gamma \pi})}\,.
\ee
The assumption that $|{\cal M}|$ is constant is not, in fact, important, as it will be discussed 
at the end of this Section. 

\subsection{Differential cross section}

\begin{figure}[!htb]
   \centering
\includegraphics[width=8cm]{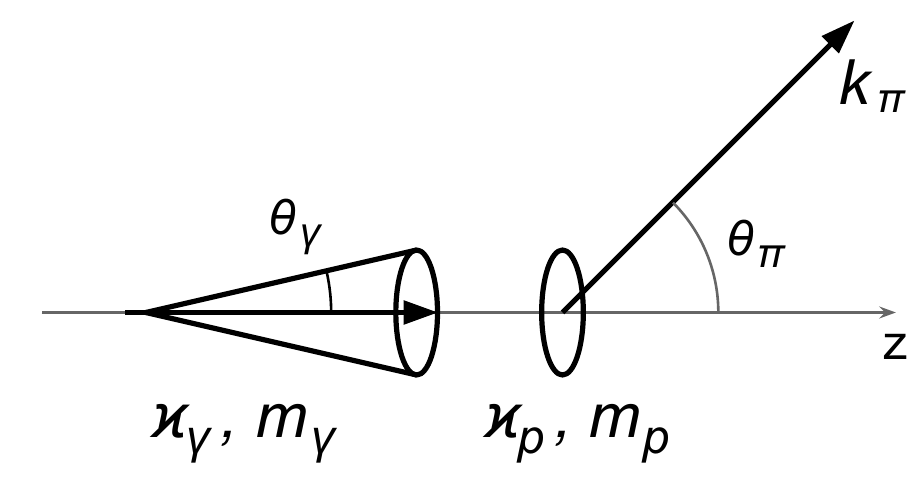}
\caption{Schematic representation of the pion photoproduction kinematics in the proton ``rest frame'' in the case
when both the photon and the initial proton are represented by the Bessel vortex states
$|\varkappa_\gamma, m_\gamma\rangle$ and $|\varkappa_p,m_p\rangle$.
}
   \label{fig-gamma-p}
\end{figure}

Our goal now is to find an efficient way to extract phase $\alpha$ (or, to be more precise,
$\alpha' \equiv d\alpha/d\theta^*_{\gamma \pi}$) 
from the vortex-beam photon-proton cross section.
To this end, we consider the double-twisted $\gamma p$ collision, in which 
both photon and proton are represented by Bessel beams 
(which will later be replaced by the transverse Bessel wave packets) 
with orbital helicities $m_\gamma$ and $m_p$.
The two twisted states are defined with respect to the same $z$ axis,
so that a longitudinal boost along this axis does not change the (transverse) vortex
structure of these states. We are allowed, therefore, we switch to the frame where the longitudinal
momentum of the proton is zero: this is the closest approximation to the rest frame one can achieve
with twisted particles. The kinematics of this process is shown in Fig.~\ref{fig-gamma-p}.

The differential cross section depends on $|F_\perp|^2$ given by (\ref{fperp2}), and in our toy model 
it is proportional to 
\be
{d\sigma \over d^2\bK} \propto 1 + \cos\left[2(m_\gamma \delta_\gamma + m_p \delta_p) + (\alpha_+ - \alpha_-)\right]\,,
\ee
where the phases $\alpha_\pm$ correspond to the two kinematical configurations shown in Fig.~\ref{fig-2-configurations}.

Suppose that the spherical angles of the initial photon are $\theta_\gamma \ll 1$, $\phi_\gamma$, 
and the angles of the final pion are $\theta_\pi$, $\phi_\pi$.
In order to calculate the phases $\alpha_\pm$, one has to take into account two kinematical effects.
First, the angle between the photon and the pion differs from $\theta_\pi$:
\be
\theta_{\gamma\pi} \approx \theta_\pi - \theta_\gamma\cos(\phi_\pi - \phi_\gamma)\,.\label{theta-gamma-pi}
\ee
Second, in the frame chosen the collision involves a proton which is not at rest but moves 
with a transverse momentum $\bk_p$, $|\bk_p|=\varkappa_p$.
Since the angle $\theta^*_{\gamma \pi}$ is defined in the proton rest frame, one needs to perform a transverse boost
with speed $v = \varkappa_p/E_p$, where $E_p \approx M_p$.
Keeping only terms which are linear in this velocity, we can relate the angles $\theta^*_{\gamma \pi}$ and $\theta_{\gamma \pi}$
with the pion energy and momentum in the lab frame ($E_\pi$, $p_\pi$) and in the proton rest frame ($E_\pi^*$, $p_\pi^*$):
$E_\pi - p_\pi \cos\theta_{\gamma \pi} = E^*_\pi - p^*_\pi\cos\theta^*_{\gamma \pi}$.
Then, we get
\be
\theta^*_{\gamma \pi} \approx \theta_{\gamma\pi} + {\varkappa_p \over M_p}
\left(1 - {\cos\theta_\pi \over \beta_\pi}\right)\cos(\phi_\pi - \phi_p)\,,
\ee
Together with (\ref{theta-gamma-pi}), this equation finally gives $\theta^*_{\gamma \pi}$.

Now, the values of $\phi_\gamma \equiv \phi_1$ and $\phi_p \equiv \phi_2$ for the two kinematical configurations 
are given by (\ref{phi1phi2}).
The difference between the two phases is then
\be
\alpha_+ - \alpha_- \approx \alpha'\, (\theta^*_{\gamma \pi\, +} - \theta^*_{\gamma \pi\, -}) = 
- 2\alpha'\, \theta_\gamma \left[1 + {E_\gamma \over M_p}\left(1 - {\cos\theta_\pi \over \beta_\pi}\right)\right] 
\sin\delta_\gamma \sin(\phi_\pi - \phi_K)\,,\label{alpha-dif}
\ee
where $M_p$ is the proton mass and we used here $\varkappa_\gamma \sin\delta_\gamma = \varkappa_p \sin\delta_p$.
Thus, we finally obtain the expression for the $\bK$-differential cross section:
\bea
{d\sigma \over d^2\bK} & \propto&  1 + \cos[2(m_\gamma \delta_\gamma + m_p \delta_p)] 
+ 2\alpha'\, \theta_\gamma \left[1 + {E_\gamma \over M_p}\left(1 - {\cos\theta_\pi \over \beta_\pi}\right)\right]\nonumber\\[2mm]
&&  \qquad\times \sin[2(m_\gamma \delta_\gamma + m_p \delta_p)]  \sin\delta_\gamma \sin(\phi_\pi - \phi_K) \,.
\label{dsigmad2K}
\eea
Note that this expression depends both on modulus $|\bK|$ via $\delta_\gamma$, $\delta_p$, and on the azimuthal angle $\phi_K$.

\subsection{Azimuthal asymmetry}

Our goal now is to extract the last term in (\ref{dsigmad2K}) which contains the desired quantity $\alpha'$.
Before we proceed, let us note the importance of using vortex beams and not just arbitrary wave packets. 
Indeed, if both initial particles did not have any phase vortex, 
then $m_\gamma = m_p = 0$, and the last term simply vanishes. In order to get the sensitivity to the phase,
one would need to consider the next term in the Taylor series, which would be quadratically suppressed 
at small $\theta_\gamma$.
{\em It is the non-zero orbital helicity that allows us to retain the term linear in small $\theta_\gamma$ and hugely boost
the sensitivity to the phase.}

Next, note that if we integrated (\ref{dsigmad2K}) over all $\phi_K$, the last term would vanish.
In order to extract it, we introduce the sine-weighted cross section
\be
\Delta \sigma = \int d^2 \bK {d\sigma \over d^2 \bK} \sin(\phi_\pi - \phi_K)\,,
\ee
to which only the last term contributes. Of course, this definition is not unique, and one could as well use the sign weight factor 
sign$[\sin(\phi_\pi - \phi_K)]$ instead of the sine itself.
The key observable is then the {\em azimuthal asymmetry}: 
\be
A = {\Delta \sigma \over \sigma}\,,\quad \mbox{where}\quad \sigma = \int d^2\bK {d\sigma \over d^2 \bK}\,.\label{asym}
\ee
To avoid misunderstanding, we remind the reader that both quantities, $\Delta \sigma$ and $\sigma$,
are still differential in $d\Omega_\pi$.

\begin{figure}[!htb]
   \centering
\includegraphics[width=10cm]{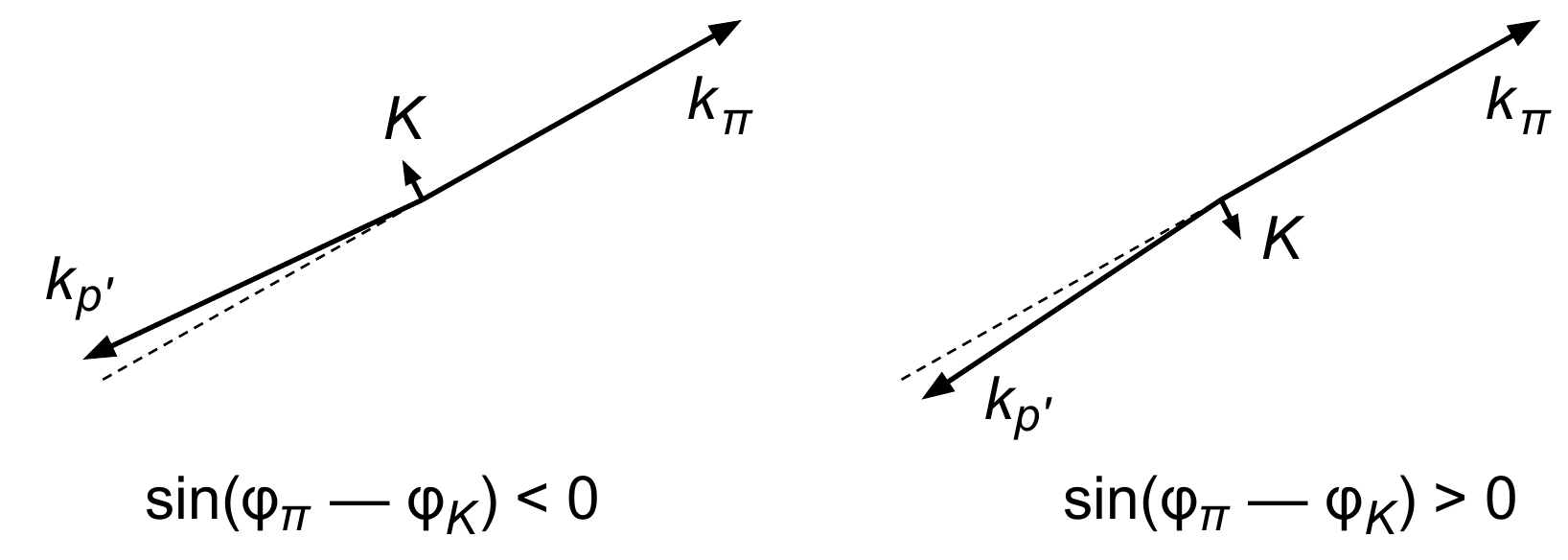}
\caption{Detecting non-collinearity of the two final transverse momenta is a prerequisite to 
measuring the azimuthal asymmetry.}
   \label{fig-non-collinear}
\end{figure}

Let us stress that in order to be able to extract the azimuthal asymmetry (\ref{asym})
from the experiment, one must be able to measure the transverse  momenta of the pion and the final proton
with a sufficient precision so that the non-collinearity of $\bk_\pi$ and $\bk_{p'}$ can be observed.
Basically, one must be able to detect the sign of $\sin(\phi_\pi - \phi_K)$, which is illustrated in Fig.~\ref{fig-non-collinear}.

Any non-zero asymmetry measured in the experiment will be an indication of a non-constant phase.
Using (\ref{dsigmad2K}), one can write the azimuthal asymmetry as 
\be
A = \alpha' \cdot P\,,
\ee
where the analyzing power $P$ can be written as
\be
P = \theta_\gamma \left[1 + {E_\gamma \over M_p}\left(1 - {\cos\theta_\pi \over \beta_\pi}\right)\right] \cdot {\cal P}\,.
\label{power}
\ee
The quantity ${\cal P}$ for pure Bessel beams can be represented as the ratio
\be
{\cal P} = \fr{\int d\bK^2\, 
\fr{\sin (2m_\gamma \delta_\gamma + 2m_p \delta_p) \sin\delta_\gamma}{\sin^2(\delta_\gamma + \delta_p)} }
{\int d \bK^2\, \fr{1 + \cos (2m_\gamma \delta_\gamma + 2m_p \delta_p)}{\sin^2(\delta_\gamma + \delta_p)}}\,. \label{calP}
\ee
If instead of pure Bessel beams one uses transverse Bessel wave packets,
then the general expressions for the azimuthal asymmetry (\ref{asym}) and the analyzing power (\ref{power})
remain the same (with replacement $\theta_\gamma \to \theta_{0\gamma} = \varkappa_{0\gamma}/E_\gamma$), 
but the expression for the quantity ${\cal P}$ becomes more complicated:
\bea
&&{\cal P} = {\int d\bK^2\, F_1 F_2 \over \int d\bK^2\, F_1^2}\,,\quad \mbox{where}\label{calP2}\\
&& F_1 = \int d\varkappa_\gamma d\varkappa_p f_\gamma(\varkappa_\gamma) f_p(\varkappa_p)\,
{\cos(m_\gamma \delta_\gamma + m_p \delta_p) \over \sqrt{\varkappa_\gamma \varkappa_p} \sin(\delta_\gamma + \delta_p)}\,,\nonumber\\[2mm]
&& F_2 = \int d\varkappa_\gamma d\varkappa_p f_\gamma(\varkappa_\gamma) f_p(\varkappa_p)\,
{\varkappa_\gamma \over \varkappa_{0\gamma}} {\sin\delta_\gamma \sin(m_\gamma \delta_\gamma + m_p \delta_p) \over 
\sqrt{\varkappa_\gamma \varkappa_p} \sin(\delta_\gamma + \delta_p)}\,.\label{F1F2}
\eea
\begin{figure}[!htb]
   \centering
\includegraphics[width=12cm]{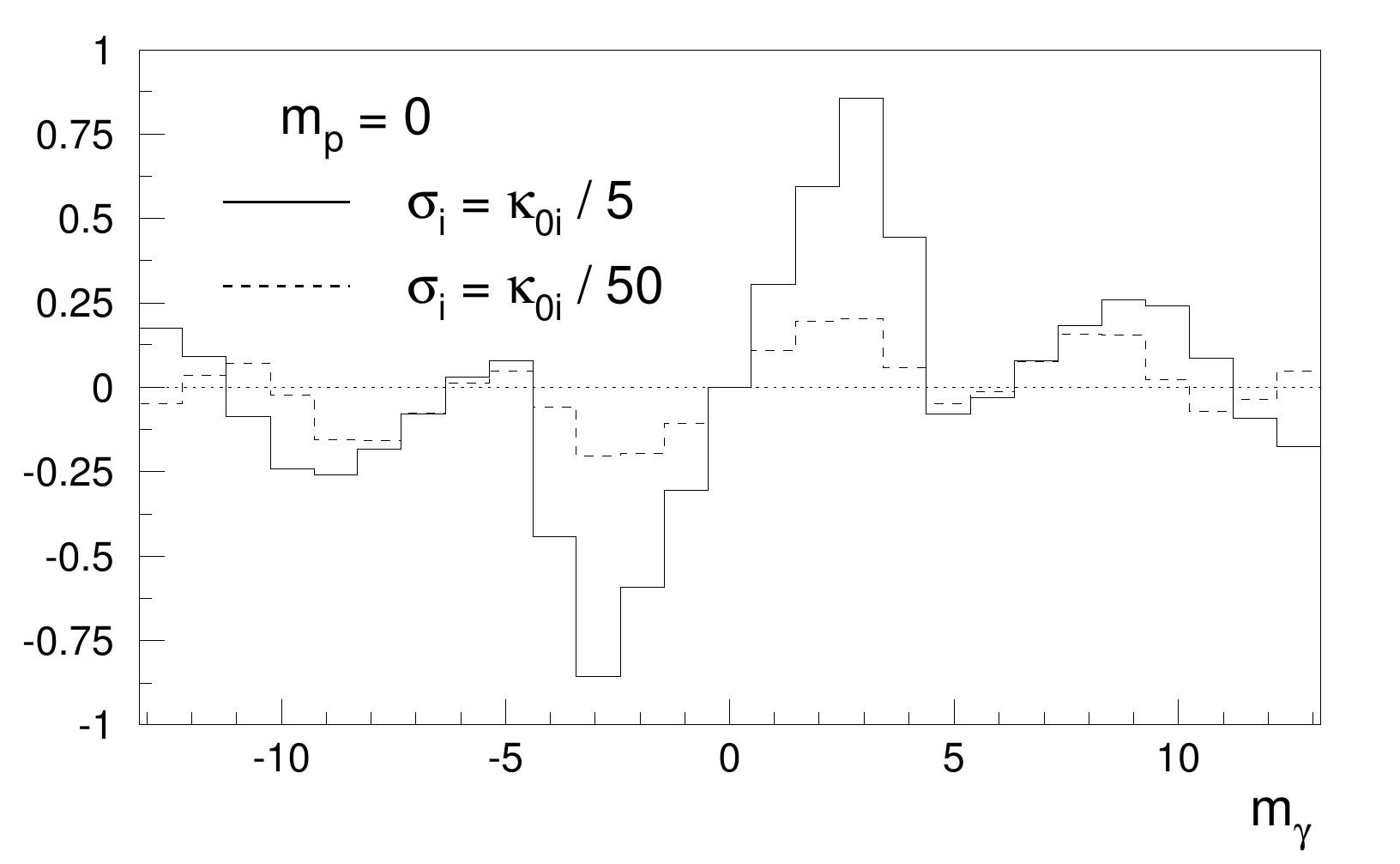}
\caption{The ratio ${\cal P}$ defined in (\ref{calP2}) for the transverse Bessel wave packets 
as a function of photon's orbital helicity $m_\gamma$; the proton's orbital helicity is fixed at $m_p = 0$.
The calculations are done for $\varkappa_\gamma = 2\varkappa_p$;
the solid and dashed lines correspond to $\sigma_i = \varkappa_{0i}/5$ and $\sigma_i = \varkappa_{0i}/50$, respectively.
}
   \label{fig-asym}
\end{figure}

Let us discuss some properties of the analyzing power $P$, (\ref{power}). First, it is linearly suppressed by the small opening angle of the photon
Bessel state, $\theta_\gamma = \varkappa_\gamma/E_\gamma \ll 1$. 
The factor in brackets is of order one; boosting it further by increasing the photon energy does not help much
because $\theta_\gamma$ will simultaneously decrease.
Finally, the last factor, ${\cal P}$, shows basically the relative strength of the sine and cosine of the angle $m_\gamma \delta_\gamma + m_p \delta_p$
averaged with the appropriate weight functions.
Of course, the denominator, as it stands in (\ref{calP}), is divergent due to the end point singularities discussed 
in Section~\ref{singularities}. On the other hand, the numerator is convergent due to the extra $\sin\delta_\gamma$
factor which turns zero at the end points. Thus, for the pure Bessel states the quantity ${\cal P}$ is 
logarithmically suppressed.
However, for the transverse Bessel wave packets this suppression is relieved.
In Fig.~\ref{fig-asym} we show this quantity for zero proton orbital helicity for a range of photon orbital helicities $m_\gamma$.
One sees that for not too large $m_\gamma$ this quantity is of order one.
In can be checked in addition that is if $\varkappa_\gamma \ll \varkappa_p$ or $\varkappa_\gamma \gg \varkappa_p$,
the quantity ${\cal P}$ is additionally suppressed; therefore, $\varkappa_\gamma \sim \varkappa_p$ will be the optimal choice.

Let us now summarize this analysis in a list of rules which one should follow in order to maximize the analyzing power.
When preparing the vortex states of the photon and proton, one should
(1) select small but non-zero orbital helicities, (2) choose $\varkappa_\gamma \sim \varkappa_p$, 
(3) choose not too small $\sigma_i/\varkappa_{0i}$.
If these criteria are met, one can expect the azimuthal asymmetry of the order 
\be
A \sim \alpha'\, \theta_\gamma\,.\label{asym-estimated}
\ee
This estimate can be safely used in qualitative discussions of the effect.

Let us finally discuss whether the assumption of the constant absolute value of the amplitude 
$|{\cal M}|$ was essential in our analysis.
In fact, it was not. Indeed, even if $|{\cal M}|$ changes with angles 
but phase $\alpha$ remains constant, then expression in the brackets in (\ref{fperp})
can be written as
\be
e^{i\alpha}\left[\cos(m_\gamma\delta_\gamma + m_p\delta_p) (|{\cal M}_+| + |{\cal M}_-|)
+ i \sin(m_\gamma\delta_\gamma + m_p\delta_p) (|{\cal M}_+| - |{\cal M}_-|)\right]\,.
\ee
The antisymmetric factor $\sin(\phi_\pi-\phi_K)$ appears only in the difference of the absolute values.
Since there is no interference between the two terms here, $|F_\perp|^2$ will contain $\sin^2(\phi_\pi-\phi_K)$, 
and the azimuthal asymmetry will be absent.
Considering transverse wave packets does not break the relative $\pi/2$ phase between these two terms
and does not change this conclusion. Therefore, even if $|{\cal M}|$ is not constant,
one can safely associate non-zero azimuthal asymmetry with varying phase.

\section{How far is this suggestion from experimental realization?}\label{section-discussion}

The measurements suggested in this paper exploit the non-plane-wave nature of the colliding beams.
Although it is not surprising that a wave packet collision is kinematically different from the plane wave
collision, we would like to stress once again the unique advantage of using vortex beams.
If both colliding beams were regular wave packets without phase vortices,
then the phase of the amplitude would be effectively averaged over a certain interval of $\theta^*_{\gamma \pi}$
with a small width $\theta_\gamma$. The difference between this average $\langle\alpha\rangle$ and its plane-wave value
would be quadratically suppressed, ${\cal O}(\theta_\gamma^2)$.
In our case, using vortex beams with a reasonably narrow $\varkappa$-distribution and measuring the azimuthal asymmetry, 
one simply removes the central value of $\alpha$ and has a very clean
phase difference between two kinematical configurations. This difference is only linearly suppressed
by the small $\theta_\gamma$. 

Next, it is true that the experiments proposed here are not yet feasible with today's technology.
As we mentioned in the introduction, so far only twisted electrons and photons with low to intermediate energies
and small values of $\varkappa$ have been created experimentally.
Though there exist suggestions of how to bring vortex beams into high-energy physics, 
their realization requires dedicated accelerator and beam physics studies.

To be more specific, the following three objectives must be achieved in order to make our proposal feasible.
\begin{itemize}
\item
Both colliding particles must be wave packets of Bessel states, and at least one of them
must contain a phase vortex, that is, must carry orbital angular momentum.
This is necessary if one wants an observable which is only linearly and not quadratically suppressed
by small $\theta_\gamma$. Arguably, the simplest way would be to create energetic twisted photons,
while for the protons it is sufficient to take a tightly focused Gaussian wave packet without any phase vortex,
which corresponds to the transverse Bessel wave packet with $m_p = 0$ and $\sigma_p \sim \varkappa_p$.
\item
The transverse momenta inside the Bessel states, $\varkappa_i$, must be sufficiently large,
namely, larger than the transverse momentum resolution of the detector.
This requirement makes it possible to detect non-collinearity of the transverse momenta of the final particles,
to determine the value of $\sin(\phi_\pi-\phi_K)$ or at least its sign and, therefore, to reconstruct
the azimuthal asymmetry.
Besides, large values of $\varkappa_\gamma$ are needed to keep the analyzing power
not too small.
\item
The initial vortex state must be prepared with non-zero but small orbital helicity, 
with $\varkappa_\gamma \sim \varkappa_p$ and not too small $\sigma_i/\varkappa_i$. 
\end{itemize}
Among these requirements, arguable the most difficult to satisfy is the second one.
For a typical hadronic experiment, it implies $\varkappa \sim {\cal O}(10-100\ \mathrm{MeV})$,
which corresponds to focusing an incoming particle to a spot of about 10 fm.
What has been achieved experimentally so far is $\varkappa \sim 10$ keV and focussing
to 1\AA. Increasing $\varkappa$ by at least three orders of magnitude seems to be a real challenge.
Nevertheless, the huge expected payoff makes it worth trying to push towards these objectives.

It is important to note that theoretical predictions of the azimuthal asymmetry (\ref{asym}) 
depend not only on the underlying microscopic physics, but also on the exact shape of the colliding
vortex beams. However detection of {\em any} non-zero azimuthal asymmetry
will signal the phase variation with the angles, as it cannot be mimicked 
by the absolute value of the amplitude.

\section{Conclusions}

In this paper we argued one can measure the angular or momentum transfer dependence
of the overall phase of the plane wave scattering amplitude by colliding vortex beams.
This measurement would be of much importance in several subfields within hadronic physics.

Since the physics of energetic vortex beams is only emerging, the proposed experiments cannot yet be realized
with today's technology, and further research in the beam physics is needed. 
We discussed the requirements that must be met in order for the measurements proposed to be feasible.
In general, we see this proposal as an example of how beneficial vortex beam collisions
can be to high-energy physics as they can probe quantities which are inaccessible in plane-wave collisions.

\section*{Acknowledgements} 

I would like to thank Jambul Gegelia for his suggestion to pay attention to the possibility 
of the phase determination in vortex beam collision.
This work was supported by the Belgian Fund F.R.S.-FNRS via the
contract of Charg\'e de recherches, and in part by grants of the Russian
Foundation for Basic Research 11-02-00242-a and NSh-3810.2010.2.

\end{document}